\documentclass[twocolumn,epsfig]{revtex4}
\usepackage{epsfig,color}
\begin{document}
\title{Effect of isospin degree of freedom on the counterbalancing of collective transverse in-plane flow}

\author{Aman D. Sood$^1$ }
\email{amandsood@gmail.com}
\address{
$^1$SUBATECH,
Laboratoire de Physique Subatomique et des
Technologies Associ\'ees \\University of Nantes - IN2P3/CNRS - Ecole des Mines
de Nantes 
4 rue Alfred Kastler, F-44072 Nantes, Cedex 03, France}
\date{\today}

\maketitle

\section*{Introduction}
Isospin degrees of freedom play an important role in heavy-ion collisions (HIC) through both nn collisions and equation of state (EOS). To access the EOS and its isospin dependence it is important to describe observables which are sensitive to isospin degree of freedom. Collective transverse in-plane flow as well as its disappearance has been found to be one such observable \cite{gautam1} where it is well known that there exists a particular incident energy called as balance energy ($E_{bal}$) at which in-plane transverse flow disappears \cite{krof89}. The disappearance of flow occurs due to the counterbalancing of attractive and repulsive interactions. In literature \cite {pak98,gautam2}the isospin dependence of collective flow as well as its disappearance has been explained to be a result of complex interplay between various reaction mechanisms, such as nn collisions, symmetry energy, surface properties of colliding nuclei and Coulomb repulsion. Here we aim to understand the effect of above mentioned mechanisms on the counterbalancing of collective flow. The present study is carried out within the framework of IQMD \cite{hart98} model. 
\section*{Results and Discussion}
We calculate $E_{bal}$ for isobaric pairs with N/Z = 1 and 1.4 throughout the mass range. 
In Figs. 1(a), (b), and (c), we display $E_{bal}$ as a function of impact parameter b
for masses 116, 192, and 240, respectively, for both the full and
reduced Coulomb. For the full Coulomb (green circles), for all the masses at
all colliding geometries, the system with higher N/Z (open symbols) has larger
$E_{bal}$ in agreement with previous studies \cite{gautam1,pak98,gautam2}. 
\begin{figure}[!t] \centering
\vskip 0.5cm
\includegraphics[angle=0,width=6cm]{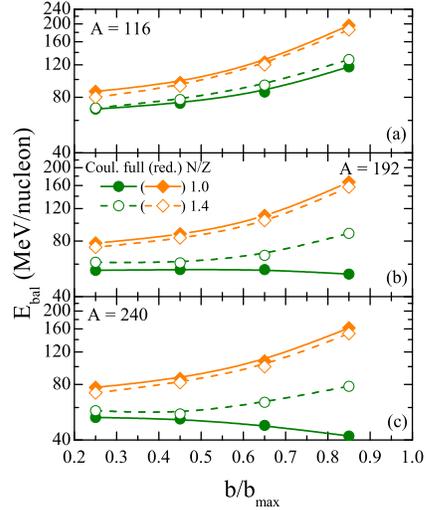}
\caption{\label{fig1} $E_{bal}$ as a function of impact parameter for different system masses. Symbols are explained in the text. Lines are only to guide the eye.}
\end{figure}
Moreover, the
difference between $E_{bal}$ for a given mass pair, increases with
the increase in colliding geometry. This is more clearly visible
in heavier masses. Also for N/Z = 1.4, $E_{bal}$ increases with the
increase in impact parameter. However for N/Z = 1 (solid symbols), the increase in $E_{bal}$ with the
impact parameter is true only for a lighter mass system such
as A = 116. For heavier masses, $E_{bal}$ in fact, begins to
decrease with the increase in impact parameter in contrast
to the previous studies \cite{gautam1,pak98}. However, when we reduce
the Coulomb [by a factor of 100 (diamonds)], we find that:
(i) Neutron-rich systems have a decreased $E_{bal}$ as com
pared to neutron-deficient systems.
\begin{figure}[!t] \centering
\vskip 0.5cm
\includegraphics[angle=0,width=6cm]{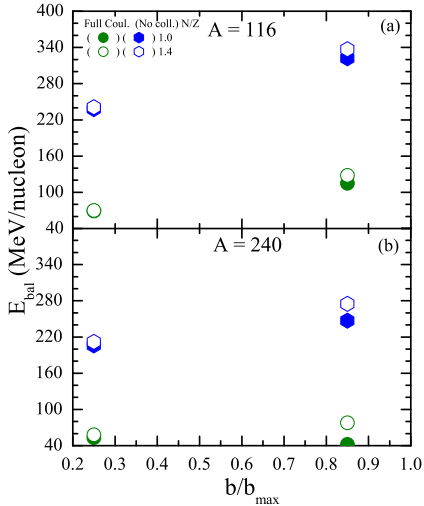}
\caption{\label{fig1} $E_{bal}$ for central and peripheral collisions with no nn collisions. Symbols are explained in the text.}
\end{figure}
This is true at all the colliding geometries
throughout the mass range. This also shows the dominance of Coulomb repulsion over symmetry energy in
isospin effects.
(ii) The difference between $E_{bal}$ for systems with different
N/Z remains almost constant as a function of colliding
geometry, which indicates that the effect of the symmetry energy is uniform throughout the range of b. This also shows that the large differences in $E_{bal}$
values for a given isobaric pair are due to the Coulomb
repulsions.
(iii) In heavier systems, at peripheral colliding geometry,
the increase in $E_{bal}$ is more in systems with N/Z =
1 compared to N/Z = 1.4, which shows the hugely
dominant role of Coulomb repulsion at high colliding
geometry and a large amount of energy is needed to counterbalance the effect of mean field. 

To understand the counterbalancing of Coulomb repulsion and mean field
in lighter and heavier systems, we switch off
the collision term and calculate the $E_{bal}$ for A = 116 and 240
at two extreme bins. The results are displayed in Fig. 2. The
hexagons represent the calculations without collisions. The
other symbols have the same meaning as in Fig. 1. We find that
at a given impact parameter $E_{bal}$ increases by a large magnitude
for both systems, which shows the importance of collisions.
The increase is of the same order for both the masses at
both impact parameter bins, indicating the same role of cross
section for both lighter and heavier masses. Moreover, the same order of increase of $E_{bal}$ at the central and peripheral colliding geometry (when we switch 
off the collisions) indicates the importance of collisions at 
high colliding geometry as well. The effect that $E_{bal}$ decreases 
with the increase in impact parameter (due to the dominance of Coulomb) for heavy mass systems with N/Z = 1 (Fig. 1, 
lower panel) does not appear here for lighter as well as heavier 
masses. This also indicates toward the increasing importance of isospin dependence of cross section with increase in impact parameter.  
Therefore in Fig. 1, the reduced Coulomb allows one 
to examine the balance of nn collisions and mean field while 
in Fig. 2 the removal of nn collisions allows one to examine 
the balance of Coulomb and mean field. 
\section*{Acknowledgments}
This work has been supported by a grant from Indo-French Centre for the Promotion of Advanced Research (IFCPAR) under project no 4104-1.


\end{document}